\documentclass[aps,prb,showpacs,superscriptaddress,twocolumn, floatfix]{revtex4}
\usepackage{amsmath}
\usepackage{amsfonts}
\usepackage{graphicx}
\usepackage{psfrag}
\begin{document}

\title{High temperature conductance of a two-dimensional superlattice \\controlled by spin-orbit interaction}
\author{P\'{e}ter F\"{o}ldi}
\email{foldi@physx.u-szeged.hu}
\affiliation{Department of Theoretical Physics, University of Szeged, Tisza Lajos k\"{o}r%
\'{u}t 84, H-6720 Szeged, Hungary}
\author{Viktor Szaszk\'{o}-Bog\'{a}r}
\affiliation{Department of Theoretical Physics, University of Szeged, Tisza Lajos k\"{o}r%
\'{u}t 84, H-6720 Szeged, Hungary}
\affiliation{Departement Fysica, Universiteit Antwerpen, Groenenborgerlaan 171, B-2020
Antwerpen, Belgium}
\author{F. M. Peeters}
\affiliation{Departement Fysica, Universiteit Antwerpen, Groenenborgerlaan 171, B-2020
Antwerpen, Belgium}

\begin{abstract}
Rashba-type spin-orbit interaction (SOI) controlled band structure of a two-dimensional superlattice allows for the modulation of the conductance of finite size devices by changing the strength of the SOI. We consider rectangular arrays and find that the temperature dependence of the conductance disappears for high temperatures, but the strength of the SOI still affects the conductance at these temperatures. The modulation effect can be seen even in the presence of strong dephasing, which can be important in case of practical applications.
\end{abstract}

\pacs{85.75.-d, 85.35.Ds, 61.50.Ah}
\maketitle

\section{Introduction}
Quantum mechanical interference effects are generally extremely sensitive to thermal fluctuations as well as to any additional dephasing mechanisms. A significant exception is the band structure of solids, which is clearly of quantum mechanical origin, but -- provided the characteristic thermal energy is well below the width of the relevant band gap -- conductance properties are still determined by the band scheme. In the following we show an example where the widths of all the band gaps can be controlled simultaneously, leading to conductance modulation also in the high temperature limit.

Periodic structures imposed on a two-dimensional electron gas where Rashba-type \cite{R60} spin-orbit interaction (SOI) is present are of fundamental importance (observable spin and flux-dependent quantum interference phenomena) and can also be promising spintronic devices.\cite{KFBP08c,A08,KKF09} The proposed arrays can be fabricated from e.g.~InAlAs/InGaAs based heterostructures \cite{KNAT02} or HgTe/HgCdTe quantum wells.\cite{KTHS06} Experiments have demonstrated that the strength of this type of SOI can be controlled by external gate voltages.\cite{NATE97,G00}
A general form of the Bloch amplitudes in a two-dimensional electron gas with Rashba-type SOI and a periodic potential has been found in Ref.~[\onlinecite{SB07}], and a spin-orbit interaction based quantum ratchet producing spin currents even in the presence of strong dissipation has also been proposed.\cite{SP07,SB08} Previous calculations\cite{FSP10} show that the band structure of these artificial crystals can be modified qualitatively, e.g., forbidden energy ranges can become allowed and vice versa simply by changing the SOI interaction strength in an experimentally achievable range. Additionally, results obtained for infinite periodic structures have strong implications on the conductance of finite arrays, e.g., forbidden bands are clearly seen as "non-conducting stripes"\cite{KFBP08b} already for relatively small arrays.

Small periodic structures like $5\times 5$\ ring arrays have recently been realized experimentally \cite{BKSN06} and have been described theoretically \cite{ZW07,KFBP08b} as well. Finite chains of quantum rings,\cite{MVP05} ladder\cite{WXE06} and diamond-like elements\cite{A08,SMC09} have been studied, as well as artificial crystal-like structures.\cite{MT09,BO10} The spin transformation properties of finite networks suggest various possible spintronic \cite{ALS02} applications as well.\cite{KFBP08c,A08,KKF09}

The characteristic energies in a superlattice with SOI controlled energy bands ("spintronic crystals") are much smaller than e.g. the usual electronic band gaps in semiconductors (i.e., meV versus eV scale). This is essentially due to the differences in the lattice constants. Clearly, the nanometer-scale translational symmetry induced subbands become important usually at low temperatures. However, as it is shown in the current paper, there is an important effect related to the simultaneous SOI-controlled modulation of the widths of the gaps in these subbands, which survives in the high temperature limit.

In the current paper, first we describe a model for the zero temperature conductance of arrays, and briefly recall the band structure related effects (Sec.~\ref{zerotempsec}). Next, in Sec.~\ref{hightempsec}, thermal fluctuations are taken into account, and a strong SOI-induced modulation of the conductance is pointed out in the high temperature limit.  In Sec.~\ref{scattsec} random scatterers are introduced with tuneable strength and we show that band gap related non-conductive stripes are surprisingly stable against this kind of dephasing. Conclusions are given in Sec.~\ref{conclsec}.

\begin{figure}[htb]
\includegraphics[width=8cm]{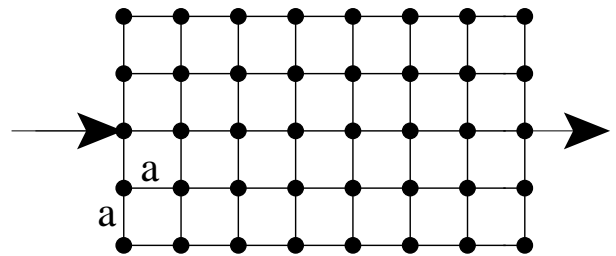}
\caption{Two-dimensional rectangular array with input and output leads. Electrons  move along the lines connecting the junctions (full circles).}
\label{latticefig}
\end{figure}

\section{Conductance at a given energy}
\label{zerotempsec}
The Hamiltonian of a narrow quantum wire in the $x$-$y$ plane with Rashba-type SOI can generally be written\cite{MMK02,MPV04}  as
\begin{equation}
H = \hbar\Omega\left[ \left( -i\frac{\partial}{\partial s}+\frac {\omega}{2\Omega} \mathbf{n}\cdot(\mathbf{\sigma}\times\mathbf{e}_z) \right) ^{2}-\frac{\omega^{2}}{4\Omega^{2}}%
\right],
\label{Ham}
\end{equation}
where the unit vector $\mathbf{n}$ points to the chosen positive direction along the wire, and we introduced the characteristic kinetic energy $\hbar\Omega=\hbar^{2}/2m^{\ast}a^{2}$ (with $a$ being the lattice constant, see Fig.~\ref{latticefig}). The strength of the SOI is given by $\omega=\alpha/a,$ where the Rashba parameter $\alpha$ \cite{MMK02} is tunable by external electric fields, and $s$ denotes the (dimensionless) length variable along the wire measured in units of $a.$ Note that -- although for $\omega=0$ Eq.~(\ref{Ham}) describes a free particle -- the complete problem is defined by the Hamiltonian \emph{and} the boundary conditions (see below) related to the geometry shown in  Fig.~\ref{latticefig}. Therefore interference phenomena that depend on the energy of the input electrons appear also in the case when the SOI vanishes. Let us also note that although in the current paper we focus on Rashba-type SOI, spin-orbit interaction related to bulk inversion asymmetry (Dresselhaus-type SOI\cite{D55}) can also play important role in the spin dynamics. For results taking both of these SOI terms into account, see e.g.~Refs.~[\onlinecite{SCL03,WV05}], while Refs.~[\onlinecite{ZFS04}] and [\onlinecite{FM07}] provide reviews discussing these interactions mainly from the spintronic point of view.

Independently from the direction of the wire, the eigenvalues of $H$ are given by
\begin{equation}
\epsilon_{\pm}=k^2 \pm \vert k \vert \frac{\omega}{\Omega},
\end{equation}
while the eigenspinor directions depend on $\delta,$ the azimuthal angle corresponding to $\mathbf{n}$:
\begin{equation}
\vert\psi\rangle_{\pm}=\frac{e^{iks}}{\sqrt{2}}\begin{pmatrix} 1\\\pm i e^{i\delta} \end{pmatrix},
\end{equation}
where units of $1/a$ have been used for the wave number $k.$
The lattice shown in Fig.~\ref{latticefig} corresponds to two Hamiltonians with orthogonal $\mathbf{n}$ vectors, and for the sake of simplicity we have chosen $\delta_1=0,$ $\delta_2=\pi/2.$
\begin{figure}[h]
\includegraphics[width=8cm]{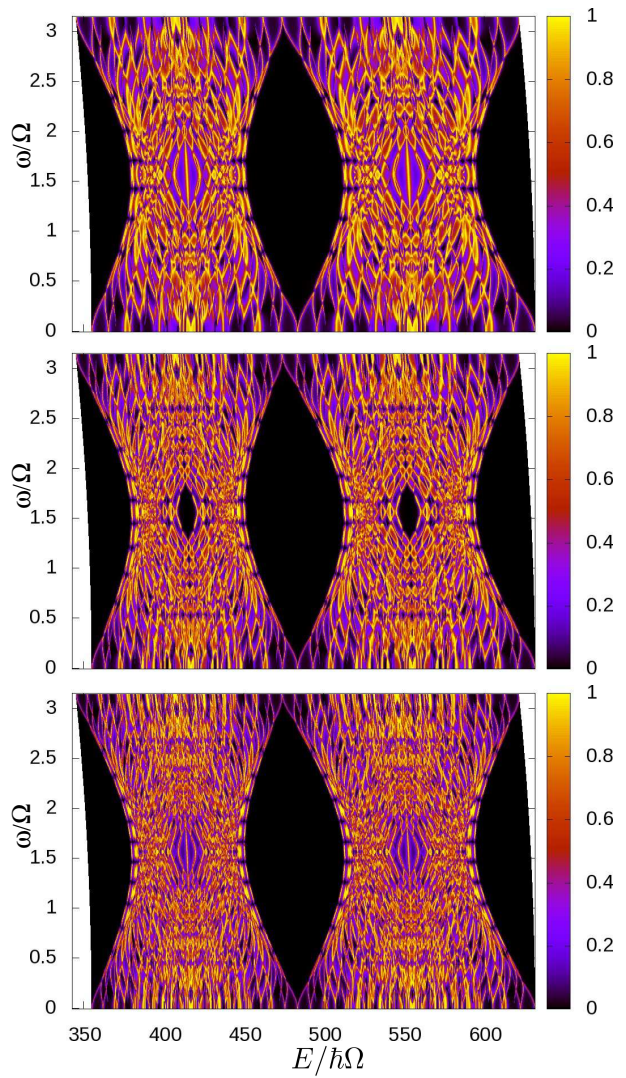}
\caption{Conductance (in units of $\frac{2e^{2}}{h}$) of rectangular arrays of different sizes (from top to bottom: $N=9,11,13$) as a function of the input energy and the strength of the SOI interaction. Note the SOI-dependent, large almond shaped minima that are directly related to the band gaps.}
\label{cond2Dfig}
\end{figure}

One of the most relevant questions related to mesoscopic arrays is the conductance as a function of the SOI strength and/or the energy of the incoming electrons. At zero temperature the latter one has zero variance, it equals the Fermi energy $E_F$ of the system. Therefore energy conservation requires finding eigenvalues of the Hamiltonian (\ref{Ham}) being equal to $E_F.$ Similarly to any other energy eigenvalue, $E_F$ is fourfold degenerate due to the two possible propagation and spin directions. In order to complete the solution of the scattering problem at a given energy, the spinor valued wave functions have to be joined together, and appropriate boundary conditions have to be implied also at the input/output junctions. We use Griffith's boundary conditions,\cite{G53} that is, the net spin current density at the junctions has to be zero and we also require the continuity of the spinor valued eigenfunctions.
Considering the physical interpretation of these boundary conditions, let us recall that the Hamiltonian leads to a continuity equation $\frac{\partial }{\partial t}\rho=\nabla \boldsymbol{j}=\frac{\partial }{\partial s }\boldsymbol{j}$ with the position dependent probability density $\rho(s)=\langle \Psi(s) |\Psi(s) \rangle.$ Inner product here is meant in the spinor sense, without integrating over spatial degrees of freedom; calculations leading to the spin current density $\boldsymbol{j}$ for a quantum ring can be found in the Appendix of Ref.~[\onlinecite{MPV04}]. The continuity equation is a local relation, its physical meaning is seen most clearly when it is integrated over a certain domain: the probability of finding the particle inside the domain changes due to the relevant currents that flow in/out through the boundaries.

Considering the input/output junctions, we assume an incoming wave and a possible reflected one at the input, while at the outputs only outgoing waves appear. The procedure above is completely analytic, in principle the transmission probabilities $T_{\uparrow}$ and $T_{\downarrow}$ for oppositely spinpolarized inputs as well as the conductance using the Landauer-B\"{u}ttiker formula\cite{D95}
\begin{equation}
G(E)=\frac{e^{2}}{h} \left[T_{\uparrow}(E) + T_{\downarrow}(E)\right]
\end{equation}
can be given in closed forms. Due to the size of the networks, it turns out to be practical to solve the related system of linear equations numerically. Let us recall that the model above assumes single mode propagation, which is a reasonable approximation for narrow conducting wires.

Contour plots of the conductance are presented in Fig.~\ref{cond2Dfig} for different $N\times N$ arrays. We can identify general, N-independent properties, most remarkably the large, almond shaped minima. These areas in the energy--SOI strength plane essentially coincide with the SOI-dependent band gaps of the corresponding lattices. According to our calculations, the band scheme for square lattices is quasiperiodic in the sense that an appropriate repetition of eight bands provides the whole band scheme.  The energies corresponding to the Bloch-wave solutions [$\exp(i \mathbf{k\cdot r}) \varphi (\mathbf{r})$ with lattice-periodic spinors $\varphi$)] can be labeled by an integer number, and $E_{n}(\mathbf{k})$ scales with the square of the band index $n.$ This fact is related to the spatial periodicity of the wave functions along the lattice, as the same phase relations at the boundaries of the unit cell can hold with e.g. $n$ and $n+1$ waves along the direction of the lattice vectors, and the dominant contribution of these solutions to the energy is proportional to $n^2/a^2$ and $(n+1)^2/a^2.$  (This quadratically scaled periodicity can be identified in the right hand side inset of Fig.~\ref{tempfig}.)
Fig.~\ref{cond2Dfig} corresponds to a complete period of the underlying band schemes, for SOI strengths that are experimentally achievable. The vertical boundaries of the relevant part of these two-dimensional plots bend slightly to the left, since increasing SOI strength shifts the band scheme downwards.

Besides strong signatures of the related band scheme, Fig.~\ref{cond2Dfig} also shows interference patterns that are less general, and their complexity increases with the size of the arrays. (See also the top panel of Fig.~\ref{deph2Dfig} for the case of $N=7.$) In the following we focus on band related phenomena, which, as we shall see later, are not only more general, but are also more stable against dephasing.

\section{SOI controlled modulation of finite temperature conductance}
\label{hightempsec}
At finite temperatures the incoming electrons are not monoenergetic, and consequently we have to average over all possible input energies. Denoting the (unnormalized) output spinor corresponding to input energy $E$ by $\left|\Psi_{out}(E)\right\rangle,$ in thermal equilibrium at temperature $T$, the output density operator is given by
\begin{equation}
\rho_{out}(T)=\int p(E,T) \left|\Psi_{out}(E)\right\rangle
\left\langle\Psi_{out}(E) \right| dE,
\label{rhoTeq}
\end{equation}
where $p(E,T)=-\frac{\partial}{\partial E}[1+\exp{(E-E_F)/k_B T}]^{-1}.$ Note that this expression corresponds to the Landauer-B\"{u}ttiker formula for the conductance at finite temperature and low bias.\cite{D95} Adding the trace of $\rho_{out}(T)$ for two oppositely polarized inputs we obtain the conductance in units of $\frac{e^{2}}{h}$.
\begin{figure}[htb]
\includegraphics[width=8cm]{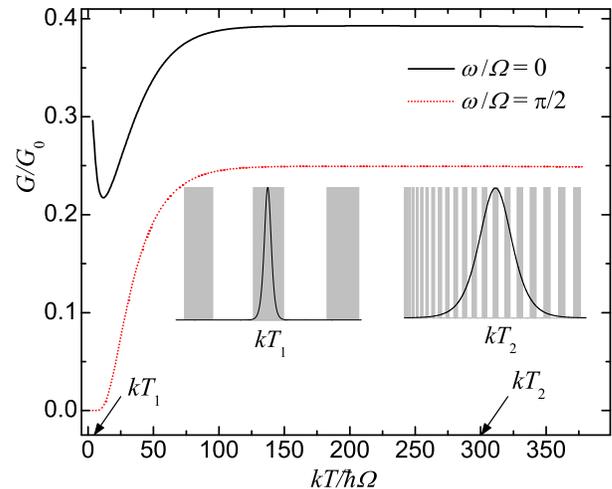}
\caption{Temperature dependent conductance (measured in units of $G_0=\frac{2e^{2}}{h}$) of a $7\times 7$ array for two different values of the SOI strength. The insets show the temperature broadened input and the band scheme (for $\omega/\Omega=\pi/2$) where gray shading corresponds to the band gaps. Note that there are no band gaps for $\omega=0.$ For InAlAs/InGaAs based heterostructures with $a=50nm$, $kT/\hbar\Omega=100$ means a temperature of 40 K.}
\label{tempfig}
\end{figure}

Conductance as a function of temperature is shown in Fig.~\ref{tempfig} for a $7\times 7$ array and for values of the SOI strength where band gaps have maximal and minimal widths (the latter is zero, see Fig.~\ref{cond2Dfig}). In order to see the most important low temperature effect as well, the size of the network is chosen such that $E_F$ is situated in the middle of a band gap for nonzero SOI. (For $\omega=0,$ when there are no band gaps at all, we use the  same value of $E_F,$ which is now obviously an allowed energy in the conduction band.) For nonzero SOI, until the width of the temperature broadened input is below that of the band gap, conductance is practically zero. (See the left hand side inset in Fig.~\ref{tempfig}.)

\begin{figure}[htb]
\includegraphics[width=8cm]{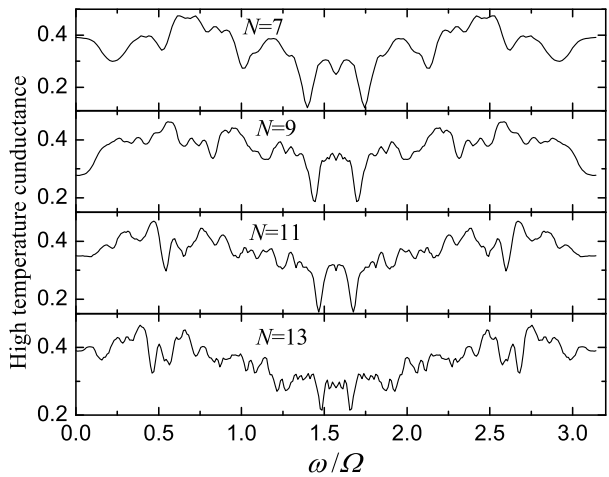}
\caption{High temperature conductance (see Fig.~\ref{tempfig}) of $N\times N$ arrays as a function of the SOI strength. (Conductance is measured in units of $G_0=\frac{2e^{2}}{h}$.)}
\label{highTfig}
\end{figure}

This low temperature effect is not particularly surprising, the most important issue here is that in contrast to smaller interference devices (like single quantum rings, where zero conductance appears only at discrete points), in the current case we have finite energy ranges with negligible transmission probabilities. Consequently, conductance modulations are still observable at finite (but low) temperatures as well.

The most remarkable feature seen in Fig.~\ref{tempfig} is the constant high temperature conductance for the two different SOI strength values. Let us note that this high temperature limit is found to be independent from the value of $E_F$, it is determined solely by the SOI strength. In the following we use the term 'high temperature conductance' for this limit, which is well defined in the framework of the model. To be concrete, we note that for InAlAs/InGaAs based heterostructures with $a=50nm$, the value $kT/\hbar\Omega=100$ corresponds to 40 K.

In order to see the physical reasons for the appearance of a constant high temperature conductance, first we recall the quasiperiodicity (as a function of energy) of the band scheme. (More precisely, energy bands are periodic as a function of $\sqrt{E},$ which is proportional to the input wave number.) For high enough temperatures (see the right hand side inset in Fig.~\ref{tempfig}), the distribution $p(E,T)$ in Eq.~(\ref{rhoTeq}) is a slowly varying function within a single period of the allowed/forbidden energy ranges. Therefore we may split the integral appearing in Eq.~(\ref{rhoTeq}) into an infinite sum over the consecutive periods in the band structure
\begin{equation}
\rho_{out}(T)\approx \sum_n p(E_n,T) \int_{E_{n1}}^{E_{n2}}\left|\Psi_{out}(E)\right\rangle
\left\langle\Psi_{out}(E) \right| dE,
\label{rhohighTeq}
\end{equation}
where $E_{n1}$ ($E_{n2}$) is the beginning (end) of the \textsl{n}th period of the band scheme. Note that the slowly varying distribution has been moved in front of the integral, and we may take $E_n=(E_{n1}+E_{n2})/2.$  This approximation is valid only for high temperatures. Conductance in this limit is not related to the fine structure of the band scheme, it is rather an overall property. Additionally, due to the periodicity of the band scheme, it is found to be sufficient to focus on a single period, evaluate the corresponding integral in the sum given by Eq.~(\ref{rhohighTeq}) and finally renormalize properly. According to our calculations, the choice of the one period long part of the band scheme to be investigated is indeed irrelevant here, and the approximation above leads to the numerically exact high temperature limit within a relative error below 5\%.

The results of the calculations based on this approximation are shown in Fig.~\ref{highTfig}. The general behavior we expect is that for zero SOI, when there are no band gaps at all, conductance is considerably higher in the high temperature limit than for cases when SOI induced band gaps are present.  As we can see, although the minima and maxima of the high temperature conductance do not correspond precisely to the widths of the band gaps (e.g.~conductance minima are not at $\omega/\Omega=\pi/2$), i.e., there are size dependent interference effects, the overall trend is the same as discussed above. As we shall see in the next section, dephasing effects average out the interference related fringes in this graph, but leave the band scheme controlled phenomena practically unchanged.

\section{Stability against dephasing}
\label{scattsec}
In a large, or even mesoscopic system transport will not be ballistic and quantum mechanical coherence of the carrier wave functions will not be maintained over the whole device. In order to give account for this issue, we introduce random, independent point-like scattering centers in the network.\cite{KFBP08b,FKP09} The considered interference phenomena in the network are spin sensitive, thus the scatterers are taken to be spin dependent. That is, we assume the presence of an additional potential
\begin{equation}
U_{scatt}^{(2)}(\mathbf{r})=\sum_n \boldsymbol{U}_n(D) \delta(\mathbf{r}-\mathbf{r}_n), \label{Scattpot2}
\end{equation}
where $\boldsymbol{U}_n(D)$ represents a $2\times 2$ diagonal matrix, with independent random diagonal elements $U_{n1}(D)$ and $U_{n2}(D).$ The Dirac delta scatterers are situated at the junctions denoted by the full circles in Fig.~{\ref{latticefig}}.
The strength of these matrix elements are random, they are taken from independent normal distributions, with zero mean and root-mean-square deviation $D.$ That is, the probability for $U_{n1}(D)$ or $U_{n2}(D)$ to have a value in a small interval around $u$ is given by $p(u)du$, where $p(u)=\exp(-u^2/2D^2)/D\sqrt{2\pi}.$ Let us note that we can interpret this model as dephasing due to random magnetic impurities at the junctions.
\begin{figure}[htb]
\includegraphics[width=8cm]{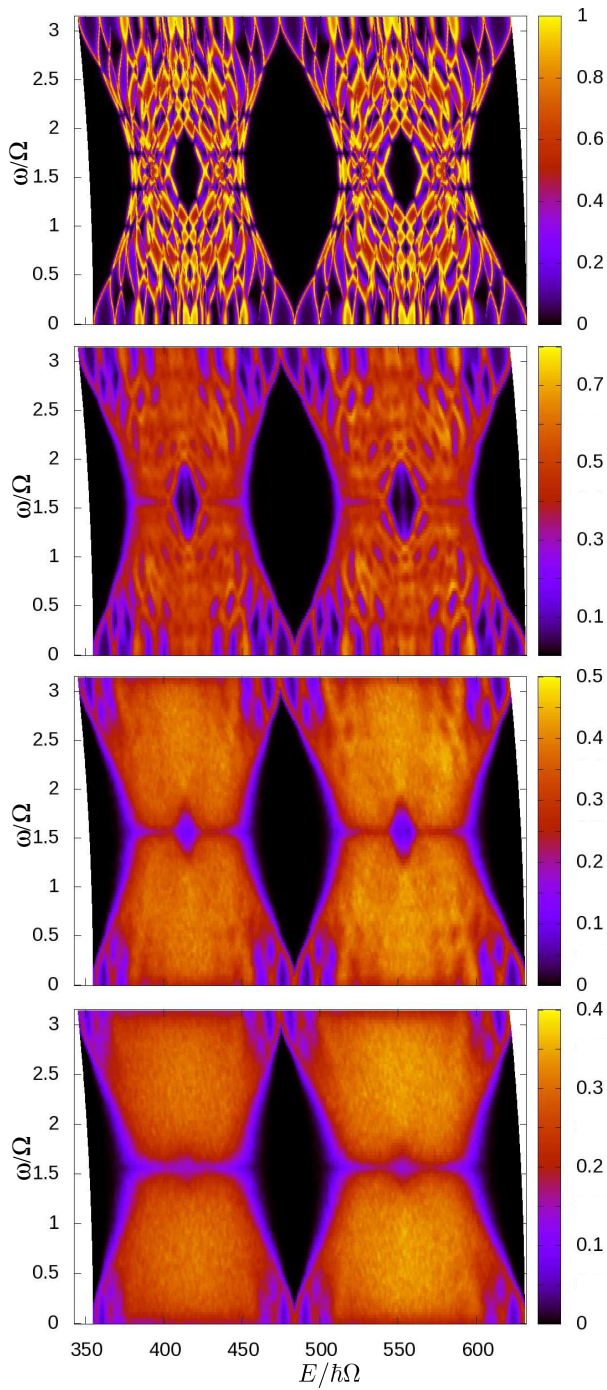}
\caption{Conductance of a $7\times 7$ rectangular array for different dephasing strengths. From top to bottom: $D/\hbar\Omega=0,10,20,30,$ and the conductance is measured in units of $G_0=\frac{2e^{2}}{h}.$}
\label{deph2Dfig}
\end{figure}

In this way, by tuning $D$ we can model weak disturbances (small $D$) as well as frequent scattering events which will completely change the character of the transport process (corresponding to large values of $D$).
Additionally, even in the presence of the spin-dependent Dirac-delta peaks, we can use Griffith's boundary conditions at the junctions, and the resulting equations are still linear.  When, after $M_c$ computational runs, the estimated output density operator $\rho_{out}(D)$ converges for a given input, we have all the possible information needed to describe the effects resulting from the disturbances characterized by the variable $D.$  Similarly to the temperature dependent case,  $\rho_{out}(D)$ is not normalized, we can consider it as a conditional density operator that describes the state of the electron if it is transmitted at all. Using the transmission probability $T,$ we have $\mathrm{Tr} [\rho_{out}(D)]=T.$
\begin{figure}[htb]
\includegraphics[width=8cm]{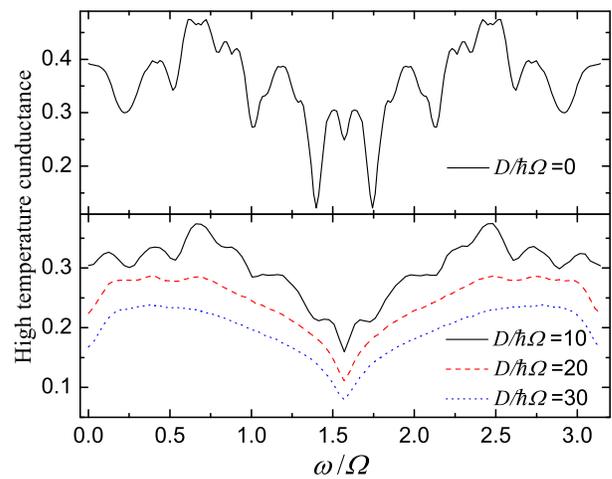}
\caption{SOI strength dependence of the high temperature conductance (in units of $G_0=\frac{2e^{2}}{h}$) of a $7\times 7$ array for different dephasing strengths.}
\label{highTdephfig}
\end{figure}

As a first application of the above method, we calculate the conductance of a given array as a function of the input energy and the SOI strength for different values of $D.$ As we can see in Fig.~\ref{deph2Dfig}, when dephasing gets stronger, the interference patterns gradually disappear, but the large, almond-shaped minima (seen already in Fig.~\ref{cond2Dfig}), that are related to the band gaps, survive. It is worth mentioning, that the average conductance decreases for larger values of $D,$ in accordance with our expectations.

Combining dephasing effects with the method described in the previous section, high temperature conductance can be calculated also in the presence of scatterers with different strengths. A representative set of results is shown in Fig.~\ref{highTdephfig}. As we can see, due to the fact that band gap related conductance minima, shown in Fig.~\ref{deph2Dfig}, are more stable against dephasing than finite size related interference patterns, high temperature conductance is still strongly modulated in the presence of moderate dephasing. In order to quantify this modulation, let us introduce the visibility
\begin{equation}
I=\frac{G_{max}-G_{min}}{G_{max}+G_{min}}
\label{viseq}
\end{equation}
that is often used also in optical applications. $G_{max}$ and $G_{min}$ denotes here the maximal and minimal SOI dependent high temperature conductance. 
\begin{figure}[htb]
\includegraphics[width=8cm]{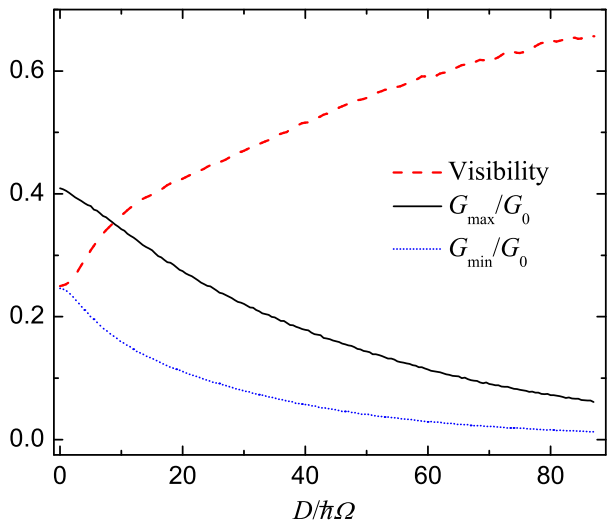}
\caption{Maximal and minimal high temperature conductance of a $7\times 7$ array for different dephasing strengths, and the corresponding visibility of the band scheme related conductance modulation. ($G_0=\frac{2e^{2}}{h}$)}
\label{visibfig}
\end{figure}
$I$ as a function of the dephasing strength is shown in Fig.~\ref{visibfig}, which can be considered as a visual summary of the current work. As we can see, SOI induced high temperature conductance modulation is still visible for strong dephasing, even when the maximal conductance drops below $25\%$ of its value at $D=0.$
Clearly, increasing visibility has little practical relevance when the conductance has essentially vanished. However, it is remarkable and promising from the viewpoint of practical applications that even in the presence of moderate dephasing and strong thermal fluctuations, the experimentally tunable SOI strength can control conductance properties.

\section{Summary and conclusions}
\label{conclsec}
In this paper we investigated high temperature conductance of two-dimensional arrays in which the propagation of the electrons is determined by the interplay of the geometry and the spin-orbit interaction (SOI). It was shown that the SOI interaction can strongly modulate the finite temperature conductance, and this effect is still present at high temperatures. We investigated how dephasing effects modify this result, and found it to be valid even when conductance is strongly suppressed due to scattering events.

\section*{Acknowledgments }
We thank M.~G.~Benedict and F.~Bartha for useful discussions. This work was supported by the
Flemish Science Foundation (FWO-Vl), the Belgian Science Policy (IAP) and the
Hungarian Scientific Research Fund (OTKA) under Contracts No.~T81364 and M045596. P.F.~was supported by a J.~Bolyai grant of the Hungarian Academy of Sciences.

\end{document}